\documentclass[a4paper,12pt]{article}

\usepackage[utf8]{inputenc}
\usepackage[dvips]{color,graphicx}
\usepackage{mathrsfs}
\usepackage{amssymb}
\usepackage{amsmath}
\usepackage{feynmp}
\usepackage{slashed}
\usepackage{latexsym}
\usepackage{graphicx}
\usepackage{verbatim}
\usepackage[normalem]{ulem}

\newcommand{\be}{\begin{equation}}
\newcommand{\ee}{\end{equation}}
\newcommand{\ba}{\begin{array}}
\newcommand{\ea}{\end{array}}
\newcommand{\bea}{\begin{eqnarray}}
\newcommand{\eea}{\end{eqnarray}}
\newcommand{\balg}{\begin{align}}
\newcommand{\ealg}{\end{align}}
\newcommand{\bit}{\begin{itemize}}
\newcommand{\eit}{\end{itemize}}
\newcommand{\trm}[1]{\textrm{#1}}
\newcommand{\mbf}[1]{\mathbf{#1}}

\newcommand{\mcl}[1]{\mathcal{#1}}
\newcommand{\mbb}[1]{\mathbb{#1}}
\newcommand{\msc}[1]{\mathscr{#1}}

\newcommand{\Mpc}{\trm{\Mpc}}
\newcommand{\yr}{\trm{\yr}}
\newcommand{\eV}{\trm{\eV}}

\newcommand{\nn}{\nonumber}

\newcommand{\Diag}[1]{\trm{diag}\{ #1 \}}

\newcommand{\model}[5]{\{m,\,k_e,\,k_\mu,\,p,\,a\} = \left\{ #1,\,#2,\,#3,\,#4,\,#5 \right\}}

\newcommand{\vtw}{\vspace{.2cm}}
\newcommand{\vth}{\vspace{.3cm}}

\newcommand{\tr}[1]{\textrm{Tr}[#1]}

\begin{document}

\title{
\Large \bf 
Discrete symmetries and model-independent patterns of lepton mixing} 
\author{
{D. Hernandez \thanks{email:
\tt dhernand@ictp.it}~~\,and
\vspace*{0.15cm} ~A. Yu. Smirnov \thanks{email:
\tt smirnov@ictp.it}
} \\
{\normalsize\em The Abdus Salam International Centre for Theoretical    
Physics} \\
{\normalsize\em Strada Costiera 11, I-34014 Trieste, Italy 
\vspace*{0.15cm}}
}
\date{}
\maketitle
\thispagestyle{empty}
\vspace{-0.8cm}
\begin{abstract}
In the context of discrete flavor symmetries, we elaborate a method that allows one to obtain relations between the mixing parameters 
in a model-independent way. Under very general conditions, we show that flavor groups of the von Dyck type, that are not necessarily finite, 
determine the absolute values of the entries of one column of the mixing matrix. 
We apply our formalism to finite subgroups of the infinite von Dyck groups, such as the modular groups, and find cases that yield an excellent agreement with the best fit values for the mixing angles. 
We explore the Klein group as the residual symmetry of the neutrino sector and explain the permutation property that appears between the elements of the mixing matrix in this case.
   
\end{abstract}

\vspace{1.cm}
\vspace{.3cm}

\newpage


\section{Introduction} \label{sec1}


During the last 10 years, the dominant approach for understanding lepton mixing has been based on the idea that it originates from a flavor symmetry $G_f$,
broken in such a way that the residual symmetries of the neutrino and charged lepton sector, $G_\nu$ and $G_l$ respectively, are different \cite{altarelli-feruglio}.
To a large extent, this approach was motivated by
the  very peculiar tri-bimaximal (TBM) mixing 
which was in agreement with experimental data \cite{tbm}. 
TBM implied the form invariance of the mass matrix - a situation  
in which the mixing matrix is determined by symmetry alone and, in particular,
does not depend on the masses of neutrinos and charged leptons.

Given the flavor group $G_f$, this generic program was realized in several ways depending on what symmetries remained in the neutrino and charged lepton sectors. The different residual symmetries in which $G_f$ could be broken were related to the Majorana nature of neutrinos and to different assignments of $G_l$-charges for the
right-handed components 
of the charged leptons. 
Taking $G_f$ as a non-abelian discrete group with a small number of elements, most possibilities were explored and shown to successfully produce interesting mixing patterns including the TBM  \cite{A4TBMmodels, S4BMmodels, S4TBMmodels,A5GRmodels}.

Yet, despite its appeal, 
this approach has certain generic problems. Since $G_f$ is broken in two different ways, there must be two distinct Higgs sectors, one for the neutrinos and another for the charged leptons, with two different vacuum alignments.
Usually some auxiliary symmetry is introduced to keep these sectors separated. 
Moreover, the necessary vacuum expectation values are achieved by introducing a number of \emph{ad hoc} assumptions and new free parameters. 
In fact, no fully convincing model has been constructed. 
In some specific realizations, the mixing is determined 
by the additional assumptions and auxiliary symmetries rather than by the original flavor symmetry.

The now established relatively large 1-3 mixing and indications of significant deviations of 2-3 mixing from maximal \cite{globalfit} imply substantial deviations from TBM mixing. This casts even more doubts on the validity of the whole approach. At the same time, it was shown that the residual symmetries can still be consistent with nonzero 1-3 mixing and nonmaximal $\theta_{23}$  \cite{ge}.

In our previous paper, we began a study whose ultimate goal is
to clarify up to which extent the flavor symmetry $G_f$ can fix the mixing 
in a model-independent way \cite{first-paper}. 
We developed an algorithm for ``symmetry building'' and used it to obtain relations between the mixing parameters  without explicit reference to any particular model. Since $G_f$ is completely broken in the whole theory, 
there are corrections to the mixing patterns derived, that do depend on the details of the specific model and cannot be obtained in our approach. Our results constitute a good approximation to those models in which those corrections are small.

More specifically, we showed under general assumptions that the flavor symmetry group, reconstructed from $G_\nu$ and $G_l$, is
of the von Dyck type. 
We derived general relations between the elements of the mixing matrix which follow directly from the group structure. 
It was shown that if a $\mbf{Z}_2$ symmetry is taken as the residual symmetry of the neutrino sector,  
one column of the mixing matrix turns out to be determined.  
This allows in particular for nonzero 1-3 mixing and a deviation of 2-3 mixing from maximal. Fixing one column of the mixing matrix produces two relations between the mixing angles and CP phase and we explored how those relations compare with  the experimental values when $G_f$ is finite.

A number of assumptions were made in \cite{first-paper} that simplified our considerations but reduced generality. In this paper, we abandon those assumptions and generalize the results of \cite{first-paper} in the following directions:

\begin{itemize}
\item We allow the charged leptons to carry arbitrary charges under $G_l$, only subject to the constraint that $G_f$ is a subgroup of $SU(3)$. This makes the results applicable to a wider class of groups. We obtain the general conditions imposed on the mixing matrix in this case. 

\item We apply the formalism to finite groups that are not of the von Dyck type but that have been used nonetheless for building models of flavor. In this paper we consider the groups $PSL(2,\mbf{Z}_7)$, $\Delta(96)$ and $\Delta(384)$ that derive from restricting the finite modular groups $\Gamma_N$ (with $N>5$) to their $SU(3)$ representations. The formalism is applicable because these groups are subgroups of infinite von Dyck groups. They were first studied systematically in \cite{toorop} where it was shown that they can lead to interesting patterns of mixing that are in agreement with recent experimental results. 
We show that the pattern of mixing derived from the corresponding von Dyck group, applies unchanged to the mixing patterns of these groups.

\item Finally, we consider the possibility that $G_\nu$ is the Klein group ($\mbf{Z}_2 \otimes \mbf{Z}_2$). We use the formalism to explain a permutation property that appears in the mixing matrices in this case. 

\end{itemize}

The paper is organized as follows. In Sec.~\ref{sec2} we present the ``symmetry building'' approach which is generalization of the approach in  our previous paper \cite{first-paper}. 
We proceed to obtain the model-independent symmetry relations between the mixing matrix elements and we show the application of this general formalism to the finite von Dyck groups.  In Sec.\ref{sec5} we generalized our formalism to finite subgroups of the infinite von Dyck groups.  Sec.~\ref{sec6} is devoted to the mixing patterns that are obtained when the complete Klein group is taken as the residual symmetry in the neutrino sector. Sec.~\ref{discussion} contains conclusions.


\section{Symmetry building} 
\label{sec2}


\subsection{Symmetries of the mass matrices and the flavor group}

We take a bottom-up approach: starting from the residual symmetries of 
the neutrino and charged lepton mass matrices, $G_\nu$ and $G_l$ with $\{G_\nu,\,G_l\} \subset G$, we reconstruct the flavor symmetry $G_f$ of the whole theory. We assume for the rest of the paper that $G_f$ is a subgroup of $SU(3)$ and that neutrinos are Majorana particles. 

In the flavor basis, the part of the leptonic Lagrangian which describes  charged current interactions and masses   
reads 
\be
\msc{L} = \frac{g}{\sqrt{2}}\bar{\ell}_L\gamma^\mu \nu_L W^+_{\mu}  + 
\bar{E}_R m_\ell \ell_L + \frac{1}{2}  \bar{\nu^c}_L M_{\nu U} \nu_L   
\label{lag} + \trm{h.c.} \, .
\ee
Here $\nu_L \equiv (\nu_e,\, \nu_\mu,\,\nu_\tau )_L^T$, $\ell_L \equiv (e,\,\mu,\,\tau)_L^T$,  $E_R \equiv (e,\,\mu,\,\tau)_R^T$ and  $m_\ell \equiv \trm{diag}\{m_e,\,m_\mu,\,m_\tau\}$. 
The neutrino mass matrix in the flavor basis,  $M_{\nu U}$, can be expressed  
via the mixing matrix $U_{PMNS}$ as 
\label{numass}
\be
M_{\nu U} = U_{PMNS}^* m_{\nu } U^\dagger_{PMNS}\, ,
\ee
where $m_{\nu } \equiv \trm{diag}\{m_1,\,m_2,\,m_3\}$ is the  
matrix of the neutrino mass eigenvalues.   
  
Let us identify the symmetries $G_\nu$ and  $G_l$ of the neutrino  and charged lepton mass terms. The generic neutrino mass matrix in Eq.~\eqref{numass} is invariant under the transformations 
\be
\nu \rightarrow S_{i U} \nu \,,\quad i = 1\,,2 \,,3 ,
\ee
where 
\be
S_{iU} = U_{PMNS}S_{i}U^\dagger_{PMNS}
\label{Siu}
\ee
and 
\be
S_1 = \Diag{1,\,-1,\,-1}\,,\quad S_2 = \Diag{-1,\,1,\,-1}\,,\quad 
S_3 = \Diag{-1,\,-1,\,1}. 
\ee
Notice that $S_3 = S_1S_2$ which implies
\be
S_{3U} = S_{1 U}S_{2U} \,.\label{S3U}
\ee
$S_{1U}$ and $S_{2U}$ commute\footnote{Notice that $S_1$ and $S_2$ 
are the symmetry transformations of the neutrino mass matrix in the mass basis.} 
and, consequently, the transformations $S_{1U}$ and $S_{2U}$ generate a Klein group, $\mbf{Z}_2\otimes \mbf{Z}_2$. Let $\mbf{Z}_{2}^{(i)}$ for $i=1,\,2$ and 3 be the $\mbf{Z}_2$-subgroup of the Klein group generated by $S_{iU}$.   
Then, $G_\nu$ can be identified either with the Klein group or with one of the $\mbf{Z}_{2i}$. The matrices $S_{iU}$ satisfy $\tr{S_{iU}} = 1$, consistent with the assumption that $G \subset SU(3)$.

\vtw
The charged lepton mass term has a full $U(1)^3$ symmetry. 
We assume that the residual symmetry in this sector is $G_l = \mbf{Z}_m$ which is a subgroup of $U(1)^3$. A representation of $G_l$ is given by the matrix $T$
such that $\ell_L$ and $E_R$ transform as \cite{first-paper}
\be
\ell \rightarrow T \ell_L \,,\quad \ell_R \rightarrow T \ell_R\,, \label{cl-transf}
\ee
where
\be
T \equiv \Diag{e^{i\phi_e},\, e^{i\phi_\mu},\, e^{i\phi_\tau}}
\label{Tdef} 
\ee
and
\be
\phi_e \equiv 2\pi \frac{k_e}{m} \,,\quad\quad \phi_\mu \equiv 2\pi \frac{k_\mu}{m} \,,\quad\quad 
\phi_\tau \equiv 2\pi \frac{k_\tau}{m} = -2\pi \frac{k_e + k_\mu}{m} \,. 
\label{phi_alpha}
\ee
$T$ satisfies the condition $T^m = \mbb{I}$. The equality 
\be
\phi_e + \phi_\mu + \phi_\tau = 0
\ee
ensures that $\trm{Det}[T] = 1$ according to the assumption that 
$G \subset SU(3)$. 

\vtw
Let us now reconstruct the full flavor group $G_f$ by taking the residual symmetries of the neutrino and charged lepton sectors as its generating elements. We consider here 
$G_\nu = \mbf{Z}_{2}^{(i)}$ and postpone to  Sec.~\ref{sec6} the analysis of the case in which $G_\nu$ is the complete Klein group.
For the charged leptons we take $G_\ell = \mbf{Z}_m$.
$G_f$ is then defined as the group formed by all $SU(3)$ matrices that can be written as a product of the generating elements of $G_\nu$ and $G_l$, $S_{iU}$ and $T$ respectively.
In particular, the matrix $W_{iU}$ defined as:
\be
W_{iU} \equiv S_{iU}T \,,
\label{rel2}
\ee 
satisfies $W_{iU} \in G$. If $G_f$ is a finite group, there must exist an integer 
$p$ such that 
\be
W^p_{iU} = (S_{iU}T)^p = \mbb{I}. 
\label{wrelation}
\ee
Thus, a minimal set of relations that can serve as a presentation of $G_f$ is given by
\be
S_{iU}^2 = T^m = W^p_{iU} = \mbb{I}.  
\label{rel1}
\ee
The three relations in Eq.~(\ref{rel1})  define the von Dyck group $D(2,m,p)$. 

In our construction it was assumed  that $G_f$ is finite. The necessary and sufficient condition for the von Dyck group $D(n,m,p)$ to be finite is
\be
\frac{1}{n} + \frac{1}{m} + \frac{1}{p} > 1 \,. 
\label{finitecond}
\ee
According to Eq.~\eqref{finitecond}, the complete list of the finite von Dyck groups is given by\footnote{In this paper we do not study the dihedral 
family $\mbf{D}_n = D(2,2,n)$. They have been less used for model building 
because they do not have irreducible representations of dimension 3. }
\be
D(2,2,n) =  \mbf{D}_n, \; 
D(2,3,3) = \mbf{A}_4,\; 
D(2,3,4) = \mbf{S}_4 ,\; 
D(2,3,5) = \mbf{A}_5\,,
\ee
which contains the most popular groups used in literature for model building.  

\vtw
For large $p$ and $m$ the condition in Eq.~\eqref{finitecond} 
is not satisfied and the von Dyck group is infinite. Finite subgroups of the infinite von Dyck groups can be generated if one imposes new relations in addition  
to those in Eqs.~\eqref{rel1}. 
These new relations should be consistent with Eq.~\eqref{wrelation}. We shall discuss those cases in Sec.~\ref{sec5}.

 



\subsection{Symmetry and relations between mixing matrix elements}
\label{sec3}


Using Eq. (\ref{Siu}), the condition in Eq.~(\ref{wrelation}) can be written 
explicitly as 
\be
W_{iU}^p = (U_{PMNS} S_i U_{PMNS}^\dagger T)^p = \mbb{I}. 
\label{main}
\ee
For $S_{i}$ and $T$ fixed, this equation can be considered 
as a condition on $U_{PMNS}$. It connects the 
group properties, determined by $S_{i}$ and $T$, with the mixing matrix. 

The explicit relations between the elements of 
 $U_{PMNS}$ imposed by Eq.~\eqref{main} 
can be obtained in the following way. 
The equation $W_{iU}^p = \mbb{I}$ implies that 
the three eigenvalues of $W_{iU}$, $\lambda_j$ ($j=1,\,2,\,3$) must satisfy~\footnote{This can be seen 
immediately since 
 $W_{iU}$ can be diagonalized by unitary matrix: 
 $W_{iU} = V^{\dagger} \hat{\lambda} V$, where 
 $\hat{\lambda} \equiv diag(\lambda_1, \lambda_2, \lambda_3)$. Then the condition 
 $W_{iU} = \mbb{I}$ is reduced to $\hat{\lambda}^p = \mbb{I}$.}
\be
\lambda_j^p = 1 \,. 
\label{p-roots}
\ee
Consider then the characteristic equation for $W_{iU}$\cite{first-paper}: 
\be
\trm{Det}[W_{iU} - \lambda \mbb{I}] = \lambda^3 - a\lambda^2 + a^*\lambda - 1 =0 \,,
\ee
where
\be
\tr{W_{iU}} = a \,.
\label{conda}
\ee
Notice that if $a$ is known, then the eigenvalues of $W_{iU}$ are completely determined. On the other hand, according to Eq.~\eqref{p-roots}, $a$ must be a sum of three $p$-th roots of unity
\be
a = \sum_{j=1}^3 \lambda_j^{(p)} \,,  ~~~~ [\lambda_j^{(p)}]^p = 1 \, .   
\label{sum}
\ee
In general, Eq.~(\ref{sum}) does not determine $a$ uniquely for any given $p$.
However, it implies that $a$ takes a discrete set of values which can be scanned systematically. For instance, we can write the possible values of $a$ for $p = 3$, 4: 
\begin{align}
p & = 3\,: \quad  a = 1 + \omega + \omega^2 = 0\,; \\
p & = 4: \quad  a = \left\{ \begin{array}{l} 
 1 \,, \\
-1+2i\,, \\
-1-2i \,,
\end{array}  \right.
\end{align}
where we have taken into account that $\prod_{j=1}^3 \lambda_j^{(p)} = 1$.
With $a$ determined as above, Eq.~\eqref{conda} is a condition on  $W_{i U}$ and, consequently, on $U_{PMNS}$.  

Thus, the condition in Eq.~(\ref{main}) is reduced to Eqs.~(\ref{conda}, 
\ref{sum}). 
Computing the trace of $W_{iU}$
and inserting it in Eq.~(\ref{conda}) we have
\be
 \sum_\alpha e^{i\phi_{\alpha}}\xi_{\alpha i} = a 
 \label{condaexp}
\ee
where
\be
\xi_{\alpha i} \equiv 2|U_{\alpha i}|^2 -1 \, \label{xii} \,.
\ee
Eq.~(\ref{condaexp}) gives two constraints on the  parameters  $|U_{\alpha i}|^2$, which correspond to the real and imaginary part of $a$. Let us introduce the notations
\be
a_R \equiv \trm{Re}\big[ a \big] \,,\quad a_I \equiv \trm{Im}\big[  a\big] \,,
\ee
and 
\be
\phi_{\alpha \beta} \equiv \phi_\alpha - \phi_\beta \,,\quad \alpha,\,\beta = e\,,\mu,\,\tau \,.
\ee
Then, using the unitarity condition, $\sum_{\alpha}|U_{\alpha i}|^2 = 1$, 
it is straightforward to solve Eq.~\eqref{condaexp} for the absolute values of the entries of the mixing matrix:
\begin{align}
|U_{ei}|^2 & =  \frac{a_R \cos \frac{\phi_e}{2} + 
\cos\frac{3\phi_e}{2} - a_I\sin \frac{\phi_e}{2} }{4\sin{\frac{\phi_{e\mu}}{2}} \sin{\frac{\phi_{\tau e}}{2}}}\,, 
\label{Uei} \\
& \nn \\
|U_{\mu i}|^2 & = \frac{a_R \cos \frac{\phi_\mu}{2} + \cos\frac{3\phi_\mu}{2} - 
a_I\sin \frac{\phi_\mu}{2} }{4\sin{\frac{\phi_{e\mu}}{2}} \sin{ \frac{\phi_{\mu\tau}}{2}}} \,,
\label{Umui} \\
& \nn\\
|U_{\tau i}|^2 & = \frac{a_R \cos{\frac{\phi_\tau}{2}} + 
\cos{ \frac{3\phi_\tau}{2}} - 
a_I\sin{ \frac{\phi_\tau}{2}}}{4\sin{\frac{\phi_{\tau e}}{2}} \sin{ \frac{\phi_{\mu\tau}}{2}}}\, . 
\label{Utaui} 
\end{align}
These equations are the central result of this paper. 
For a specific $S_i$ and a selected set of  parameters 
$\{m,\, k_e,\, k_\mu,\, p,\, a\}$,
they give the absolute values of the $i$-th column 
of $U_{PMNS}$:  
\be
\mcl{V}_i^T (m, k_e, k_\mu, p, a) \equiv (|U_{ei}|^2, ~|U_{\mu i}|^2, ~ |U_{\tau i}|^2 ). 
\ee 
We call the five parameters $\{m,\,k_e,\,k_\mu,\,p, \, a \}$ the \emph{symmetry assignment}. The first parameter is the order of the residual symmetry $G_l$ while the second and third ones are the lepton charges under this subgroup. The fourth and fifth parameters are properties of $W_{iU}$. They link $G_\nu$ and $G_l$ and give the restrictions on the mixing matrix. The von Dyck group is defined by $m$ and $p$ while $k_e$, $k_\mu$ and $a$ specify its three-dimensional representation.
 
\vtw
It can be easily shown that a permutation of $m$ and $p$ corresponds to the same von Dyck group. 
Indeed, taking into account that $S_{iU}^2 = \mbb{I}$, one has
\be
S_{iU}W_{iU} = T\,. 
\label{SW=T}
\ee
Substituting $T$ from this equation in Eq.~\eqref{rel1} 
we obtain
\be
S_{iU}^2 = W^p_{iU} = (S_{iU}W_{iU})^m =  \mbb{I}.  
\label{rel3}
\ee
These relations define the von Dyck group $D(2,\,p,\,m)$ which proves that  $D(2,\,m,\,p) = D(2,\,p,\,m)$. 


The presentation in Eq.~(\ref{rel3}) has the same form of Eq.~\eqref{rel1} with the exchanges $W_{iU} \leftrightarrow T$ and $m \leftrightarrow p$. This implies that $W_{iU}$ can be taken as the generator of $G_l$
instead of $T$. In that case, the last equality in Eq.~(\ref{rel3}) imposes relations between the mixing matrix elements in the same way as the equality $(S_{iU} T)^p = \mbb{I}$ did before.  

However, since $W_{iU}$ is not diagonal, the presentation in Eq.~(\ref{rel3}) is in a basis that is not the flavor basis.  To get the mixing matrix in the flavor basis we apply the unitary transformation $V$ that diagonalizes $W_{iU}$ on the elements of $G_f$. Hence, we define  
\be
T' \equiv V^\dagger W_{iU}V  \,, \quad
S'_{iU} \equiv V^\dagger S_{iU}V\, ,  
\label{Vtransf}
\ee
where $T'$ is diagonal. The new generator of $G_l$ 
can be written as 
\be
T' = \Diag{e^{i\phi'_e},\, e^{i\phi'_\mu},\, e^{i\phi'_\tau}}\label{T'def} 
\ee
with 
\be
\phi'_e \equiv 2\pi \frac{k'_e}{p} \,,\quad \phi'_\mu \equiv 2\pi \frac{k'_\mu}{p} \,,
\quad \phi'_\tau \equiv 2\pi \frac{k'_\tau}{p} =  -2\pi \frac{k'_e +k'_\mu}{p} \,. 
\label{phi'_alpha}
\ee

The group presentation does not change  when written in terms of $S'$ and $T'$ so we have
\be
S'^2_{iU} = T'^p = (W'_{iU})^m = \mbb{I}\, , 
\label{S'T'} 
\ee
where 
\be
W'_{iU} \equiv S'_{iU} T' = V^\dagger T V.     
\label{defwprime}
\ee
Following our derivation of Eq.~(\ref{conda}), we obtain that the last equality in Eq.~(\ref{S'T'}) leads to a condition for the elements of the new mixing matrix $U'_{PMNS}$ that we can write as:
\be
\tr{W'_{iU}} = a' = \tr{T} .
\label{eqtr}
\ee
Here the last equality follows from Eq.~(\ref{defwprime}). Thus, $a'$ can be  obtained immediately by computing the trace of the original matrix $T$. 

We call the symmetry assignment $\{p,\,k'_e,\,k'_\mu,\,m,\,a'\}$, $mp$-\emph{permuted} with respect to  $\{m,\,k_e,\,k_\mu,\,p, \, a \}$. The new mixing matrix $U'_{PMNS}$ can be obtained from the matrix  of the original assignment,  $U_{PMNS}$, by noticing that 
\be
S'_{iU} = V^\dagger S_{iU} V = 
V^\dagger U'_{PMNS} S_i {U'_{PMNS}}^{\dagger} V , 
\ee
and therefore
\be
U'_{PMNS} = V^\dagger U_{PMNS} \,. 
\label{U'PMNS} 
\ee


\subsection{Applications. Finite von Dyck groups}
\label{sec4}


As it was established in the previous section, the procedure of finding relations between the mixing matrix elements, given a specific symmetry assignment, consists of the following steps:

\begin{enumerate}

\item
Using Eq. (\ref{sum}), find $a = a\big(\lambda_i^{(p)} \big)$. 

\item From $m$, $k_e$, $k_\mu$, determine the phases $\phi_\alpha = \phi_\alpha (m, k_e, k_\mu)$ and  $\phi_{\alpha \beta} = \phi_{\alpha \beta}(m, k_e, k_\mu)$ according to Eq.~(\ref{phi_alpha}). 

\item Using $a_R$, $a_I$, $\phi_\alpha$ and  $\phi_{\alpha \beta}$, determine  $|U_{\alpha i}|^2$ from Eqs.~(\ref{Uei}-\ref{Utaui}).  

\item Substituting the standard parameterization of $U_{PMNS}$ in Eqs.~(\ref{Uei}-\ref{Utaui}), obtain the two conditions for the three mixing angles and CP phase. 

\end{enumerate}
We apply this procedure in what follows. 

\vth
Eqs.~(\ref{Uei}-\ref{Utaui}) generalize the results obtained in \cite{first-paper} where  a vanishing value for one of $k_e$, $k_\mu$ or $k_\tau$ was taken. This corresponds to $a_I=0$ and real values of both $\tr{W}$ and $\tr{T}$. In this case  $T = T_\alpha$, with $\alpha = e$, $\mu$ or $\tau$  has the diagonal elements
\be
T_{\alpha \alpha} = 1, ~~~T_{\beta \beta} = T_{\gamma \gamma}^* = e^{2\pi i k/m}, ~~~~\beta \neq \gamma \neq \label{Talpha}
\alpha. 
\ee
For instance, $T_e  = \trm{diag} \{1,\, e^{2\pi i k/m},\,e^{-2\pi i k/m}\}$, etc.. 
For each of these cases Eqs.~(\ref{Uei}-\ref{Utaui}) reduce to 
\be
|U_{\alpha i}|^2  = \eta, ~~~~  
|U_{\beta i}|^2   = |U_{\gamma i}|^2\, ~~~   \beta,\,\gamma \neq \alpha, 
\label{main1}
\ee
where 
\be
\eta  \equiv \frac{1+a}{4 \sin^2\left( \frac{\pi k}{m} \right)}. 
\label{eta}
\ee
In  \cite{first-paper} several cases were analized in which  Eqs.~(\ref{main1}, \ref{eta}) lead to values for the mixing angles that are compatible with the experimental results. For instance, taking  the symmetry  assignment $\model{3}{0}{1}{4}{1}$, that corresponds to taking $T_e$ in Eq.~\eqref{Talpha}, we obtain the column of mixing parameters 
\be
\mcl{V}_i^T =  \left( \frac{2}{3},\, \, \frac{1}{6},\, \, \frac{1}{6} \right) \, .  
\label{TM1}
\ee
For $\model{3}{0}{1}{3}{0}$ the column equals 
\be
\mcl{V}_i^T = \left( \frac{1}{3},\, \, \frac{1}{3},\, \, \frac{1}{3} \right)\,. 
\label{TM2}
\ee
Both predictions in  Eq.~\eqref{TM1} and Eq.~\eqref{TM2} are in agreement with non-zero $\theta_{13}$. The case of  Eq.~\eqref{TM1}  applied to $i=1$, is known in the literature as the Trimaximal 1 (TM1) mixing pattern while  the second case, Eq.~\eqref{TM2}  for $i=2$, is the Trimaximal 2 (TM2) pattern \cite{tm}. The derivation presented here shows that TM1  and TM2 are specific cases of the general formulae in Eqs.~(\ref{Uei}-\ref{Utaui}) that, in turn, can be derived directly from the group structure. 

\vth
Let us consider the choice of parameters $\model{4}{3}{0}{3}{0}$ 
that has not been explored in previous works.  $T_\mu$ is given by $\Diag{-i,\,1,\,i}$ and the corresponding group is $\mbf{S}_4$. In this case  $|U_{\mu 2}|^2 = \frac{1}{2}$ and the second column fixed at 
\be
\mcl{V}_2^T = \left(\frac{1}{4} , \, \, \frac{1}{2}, \, \, \frac{1}{4} \right). 
\label{newcase}
\ee 
Using the standard parametrization, Eqs.~\eqref{newcase} yield the following relations between the mixing angles
\begin{align}
\sin^2\theta_{12} \cos^2\theta_{13} & = \frac{1}{4}\,, 
\label{s12m4k1} \\
\cos^2\theta_{12} \cos^2\theta_{23} & + \sin^2\theta_{12}\sin^2\theta_{23}\sin^2\theta_{13} - \nn \\
& \quad\quad\quad - 2\cos\theta_{12}\sin\theta_{12}\cos\theta_{23}\sin\theta_{23}\sin\theta_{13}\cos\delta = \frac{1}{2} \,.
\label{s23m4k1}
\end{align}
In the limit of $\theta_{13} \rightarrow 0$ these equalities give  $\sin^2\theta_{12} = 1/4$ and $\sin^2\theta_{23} = 1/3$. 
\begin{figure}[t]
\begin{center}
\includegraphics[width=7.5cm]{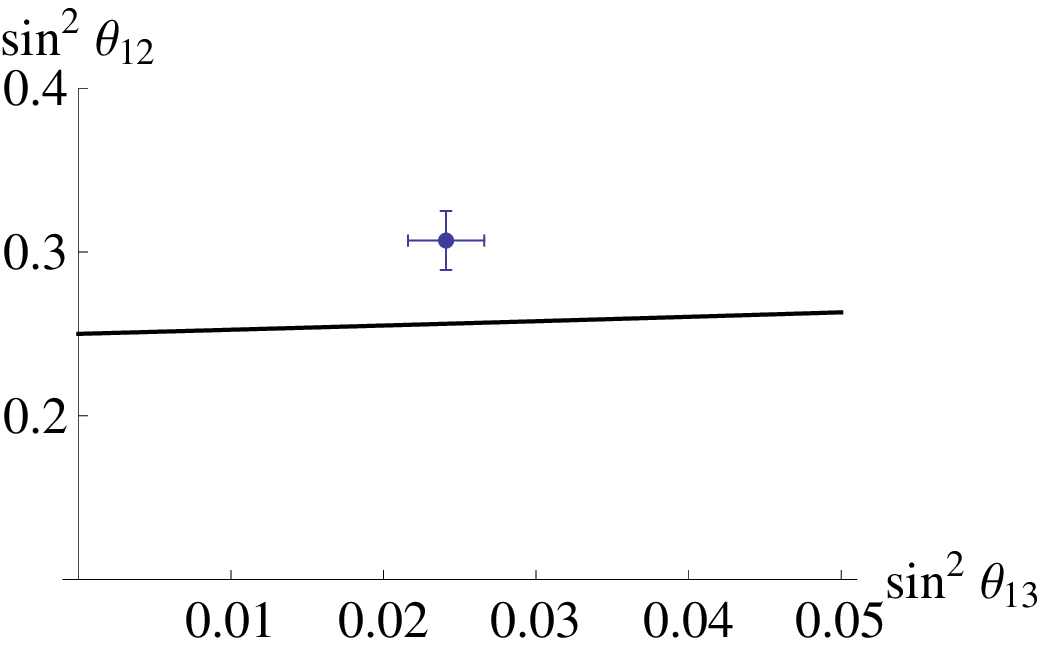}
\includegraphics[width=7.5cm]{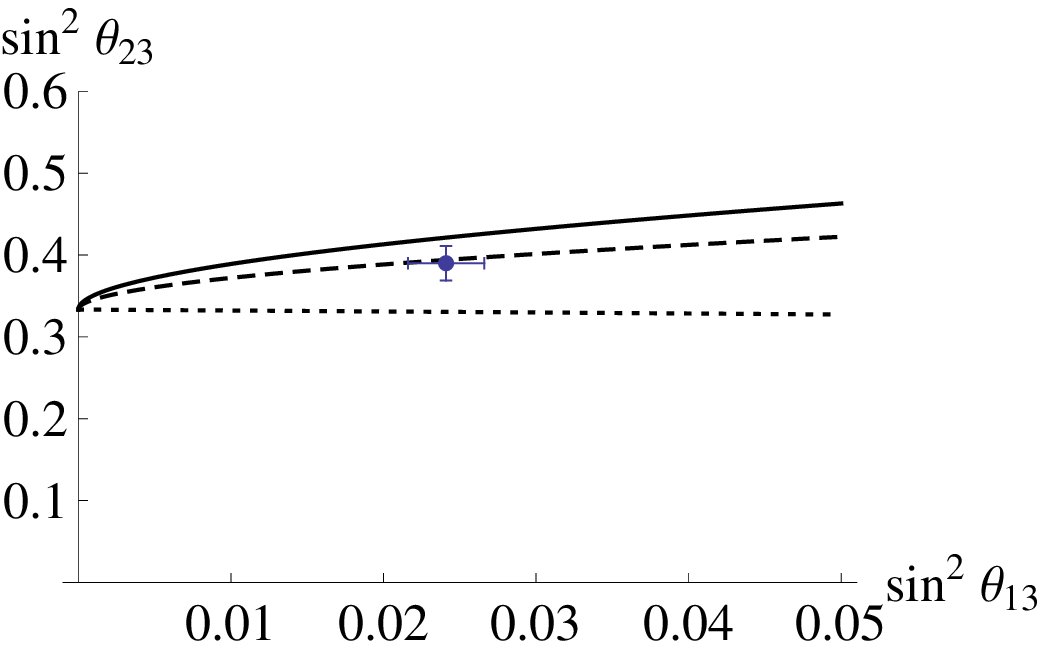}
\caption{
$\sin^2\theta_{12}$ (left panel) and  $\sin^2\theta_{23}$ (right panel) as  functions of $\sin^2\theta_{13}$ for the assignment  $\model{4}{3}{0}{3}{0}$. 
The curves  in the right panel correspond to $\delta = \pi$ (thick), $3\pi/4$ 
(dashed),  $\pi/2$ (dotted).}
\label{newfig1}
\end{center}
\end{figure}
In Fig.~\ref{newfig1} we plot the values of $\sin^2\theta_{12}$ 
and $\sin^2\theta_{23}$ that result from Eqs.~\eqref{s12m4k1} 
and \eqref{s23m4k1}. The prediction for $\sin^2\theta_{12}$ 
is more than 3$\sigma$ away from the best fit value, so that this case could  
correspond to a first order approximation and large corrections are required.

\vth
For the case of finite 
von Dyck groups, we performed a scan over all possible assignments 
in Eqs.~(\ref{Uei}-\ref{Utaui}). We find that the only assignments  compatible with the requirement that $0 \leq |U_{\alpha i}|^2 \leq 1$ are precisely the ones that correspond to real  $\trm{Tr}[W]$ and $\trm{Tr}[T]$. In other words, the relations  found in \cite{first-paper} and in this section describe all the possible mixing patterns that can be derived  from the finite von Dyck groups in this framework.


\section{Finite subgroups of the infinite von Dyck groups} 
\label{sec5}


\subsection{General considerations}


In this section we extend the analysis of the previous sections to finite subgroups of infinite von Dyck groups. As examples, we will consider 
the groups $PSL(2,\, \mbf{Z}_7)$, $\Delta(96)$ and $\Delta(384)$ explored in  
\cite{toorop}.    
These groups can be generated by the matrices $S_{iU}$ and $T$ that satisfy 
the von Dyck relations,  Eq.~\eqref{rel1}, 
with $\{p,\,m\} = \{7, 3\}$ or $\{7,\,4\}$ for $PSL(2,\, \mbf{Z}_7)$, 
 $\{p,\,m\} = \{8, 3\}$ for $\Delta(96)$ and $\{p,\,m\} = \{16, 3\}$ for $\Delta(384)$. 
In order to render the group finite, a relation of the form
\be
X^q  = \mbb{I} \,,\quad X \equiv S_{iU}T^{-1}S_{iU}T  \,,
\label{relX} \,
\ee
where $q$ is an integer, is added to Eq.~(\ref{rel1}).  
For the case of the finite von Dyck groups discussed in the previous section, Eq.~(\ref{relX}) is redundant after Eq.~(\ref{rel1}). In the case of infinite groups, the equality in Eq.~(\ref{relX}) is in general invalid as we will see below.  Hence, it imposes additional restriction on the parameters of the group. 

Notice that $S_{iU}X^\dagger S_{iU} = X$ and, taking into account that $S = S^{-1}$, we find that $X$ and $X^\dagger$ are similar matrices with the same eigenvalues. Therefore, if $\lambda$ is an eigenvalue of $X$, $\lambda^*$ must also be. Since $X$ is a $3\times 3$ matrix, we conclude that $X$ must have one eigenvalue equal to 1 and that $\tr{X}$ must be a real number.

\vtw
We will use the same method as in  in Secs. 2 and  3   
to analize the relation in Eq.~\eqref{relX}.  
In particular, from  the equality in Eq.~\eqref{relX}  we obtain 
\be
\tr{X} = x,   
\label{condx}
\ee
where $x$ is the sum of three $q-$th roots of unity 
\be
x = \sum_{j=1}^3 \lambda_j^{(q)} \,,  ~~~~ [\lambda_j^{(q)}]^q = 1 \, , 
\label{sumx}
\ee
and it takes a discrete set of values. 
The trace in Eq. (\ref{condx}) equals
\begin{align}
\tr{X} & = \xi_{ei}^2 + \xi_{\mu i}^2 + \xi_{\tau i}^2 + 
8\left[ \cos \phi_{e\mu}|U_{e i}|^2|U_{\mu i}|^2  + 
\cos\phi_{e\tau }|U_{ei}|^2|U_{\tau i}|^2 \right. \nn \\ 
& \left. \quad\quad\quad\quad\quad \quad\quad\quad\quad + 
\cos\phi_{\mu\tau}|U_{\mu i}|^2|U_{\tau i}|^2 \right] 
\nn \\
& = |\tr{W}|^2 - 2  \left[ \cos\phi_{e\mu} \xi_{\tau i} + 
\cos\phi_{\mu\tau} \xi_{ei} + \cos\phi_{\tau e}\xi_{\mu i} \right]  
\label{condX} \,.
\end{align}
Defining
\be
A = 2 + e^{i\phi_{e\mu}} + e^{-i\phi_{e\mu}} + e^{i\phi_{\mu\tau}} + e^{-i\phi_{\mu\tau}} + e^{i\phi_{\tau e}} + e^{-i\phi_{\tau e}}\,,
\label{a-phases}
\ee
we can rewrite Eq.~\eqref{condX} as 
\be
\tr{X} =  |\tr{W_{iU}}|^2 + \tr{W_{iU}}^\dagger \tr{T} + \tr{W_{iU}}\tr{T}^\dagger + A \,. 
\label{finalcondX}
\ee
Note that the RHS in Eq.~\eqref{finalcondX} is invariant under 
the permutation $T \leftrightarrow T^\dagger$. That is, taking $\{S_{iU} ,\,T\}$ or $\{S_{iU},\, T^\dagger\}$ 
as generators leads to the same $\tr{X}$. 

Substituting 
Eq.~(\ref{conda})  in  Eq.~\eqref{finalcondX} we obtain 
\be
x = |a|^2 + a\tr{T}^\dagger + a^\dagger \tr{T} + A 
\label{conda'exp} \,.
\ee
This equality does not depend on $U_{PMNS}$ and, therefore, the additional relation in Eq.~(\ref{relX}) - or Eq.~(\ref{condx}) - does not add any new constraint to the mixing parameters.  
Instead, Eq.~(\ref{conda'exp}) 
puts a condition that the eigenvalues of the matrix $T$ must satisfy. 


For some particular values of $p$ and $a$, Eq.~\eqref{conda'exp} acquires a particularly simple form. For $p = 3$ and  $a = 0$, Eq.~\eqref{finalcondX} reduces to 
\be
x = A \, .  
\label{trXp=3}
\ee
For $p=4$ and $a = 1$, Eq.~\eqref{finalcondX} reads 
\be
x = 1 + \tr{T} + \tr{T}^\dagger + A \,.   
\label{trXp=4}  
\ee
Notice that  Eq.~\eqref{trXp=4} is satisfied, in particular,  when $x = A$ and $\trm{Re} \big( \tr{T} \big) = - 1/2$. If $A$ is found from Eqs. (\ref{trXp=3}) or  (\ref{trXp=4}) or in general from (\ref{conda'exp}), then the Eq. (\ref{a-phases}) with known $A$ can be used to obtain $k_e$ and $k_\mu$.

As in the case of finite von Dyck groups, the $mp$-permuted solution exists. Indeed, since $T = S_{iU}W_{iU}$, the matrix $X$ can be rewritten in terms of $S_{iU}$ and $W_{iU}$ as
\be
X = S_{iU}(T^{-1}S_{iU})S_{iU}(S_{iU}T) = S_{iU}W_{iU}^{-1}S_{iU}W_{iU} \,.
\label{XSW}
\ee
That is, the element $X$ has the same form when written in terms of $S_{iU}$ and $T$ and in terms of $S_{iU}$ and $W_{iU}$ so that the elements $W_{iU}$ and $T$ can exchange roles. Using  $S'_{iU}$ and $T'$ as given in  Eq.~(\ref{Vtransf}),   
we obtain 
\be
S'^2_{iU} = T'^p = (W'_{iU})^m = ( S'_{iU}T'^{-1}S'_{iU}T')^q = \mbb{I} \,.
\ee
Thus,  $S'_{iU}$ and $T'$ generate the same group $G_f$. 
Imposing that $T'$ is the residual symmetry of the lepton sector leads 
to a different mixing matrix $U'_{PMNS}$ given by Eq.~\eqref{U'PMNS}. 
Similarly to Eq.~(\ref{eqtr})
the values of $k_e$ and $k_\mu$ which determine $T$ also    
determine the eigenvalues of $W'_{iU}$. 


\subsection{Symmetry relations from $PSL(2,\,\mbf{Z}_7)$}


The group $PSL(2,\,\mbf{Z}_7)$ is a subgroup of the infinite von Dyck group with $p = 3$ and $m=7$,  so that the group presentation is
\be
S_{iU}^2 = T^7 = (S_{iU}T)^3 = \mbb{I} \, .
\label{prepsl}
\ee
The additional relation, Eq.~\eqref{relX},
reads 
\be
X^4 = \mbb{I} 
\label{X^4}\,.
\ee
\begin{figure}[ht]
\begin{center}
\includegraphics[width=7.5cm]{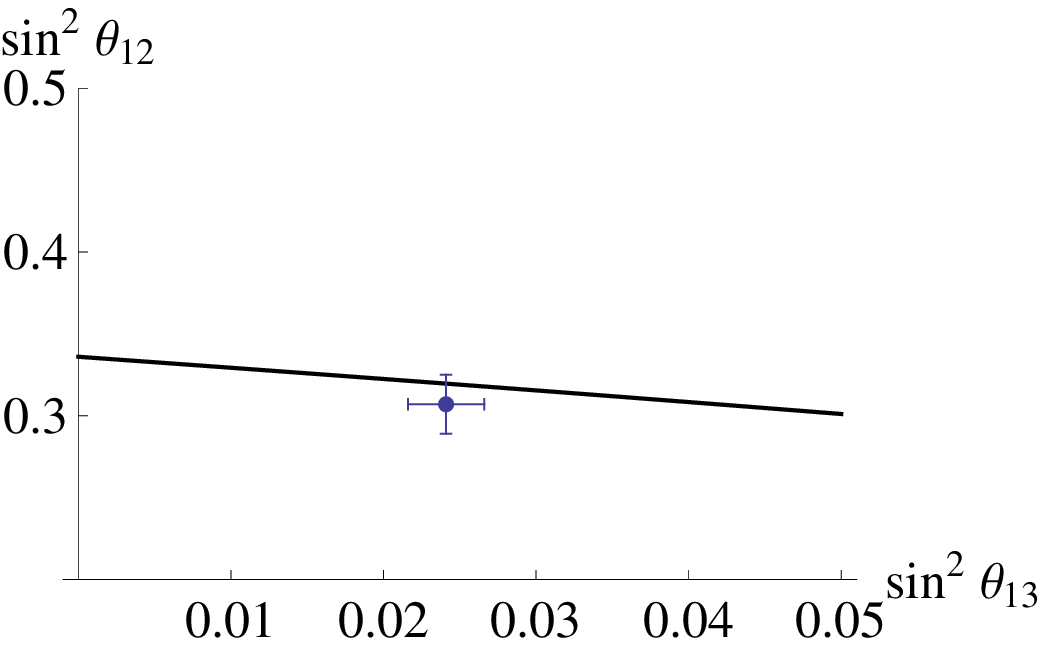}
\includegraphics[width=7.5cm]{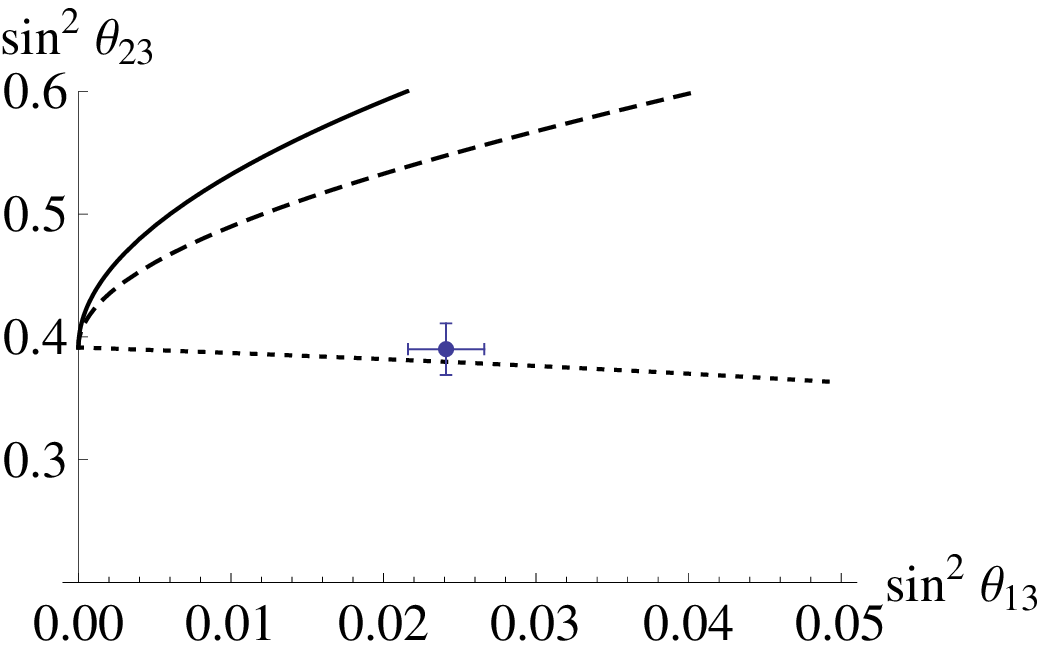}
\includegraphics[width=7.5cm]{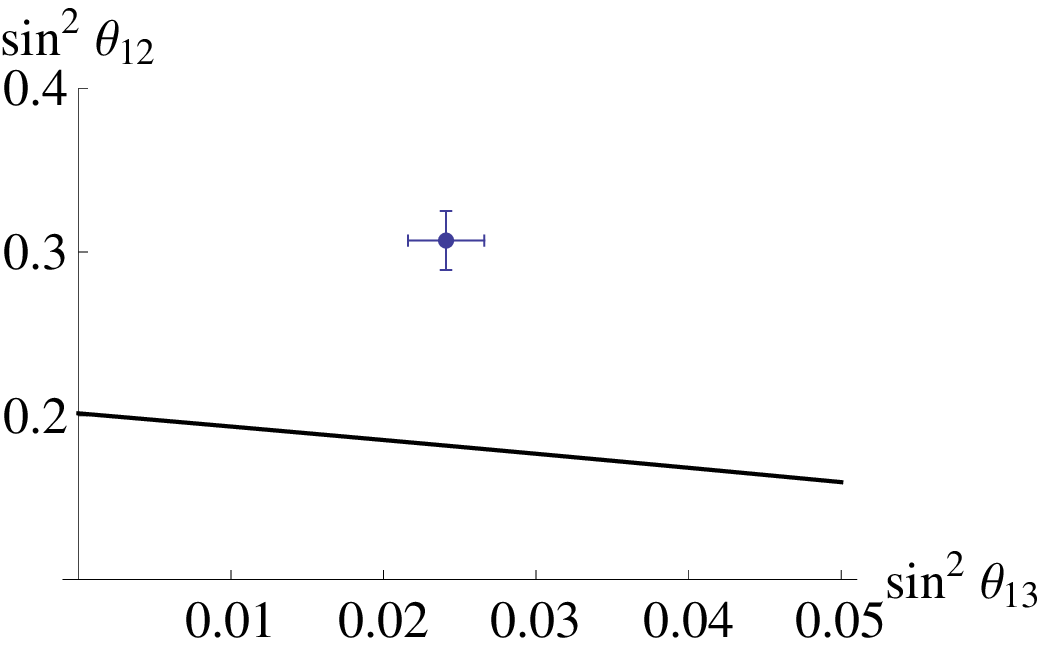}
\includegraphics[width=7.5cm]{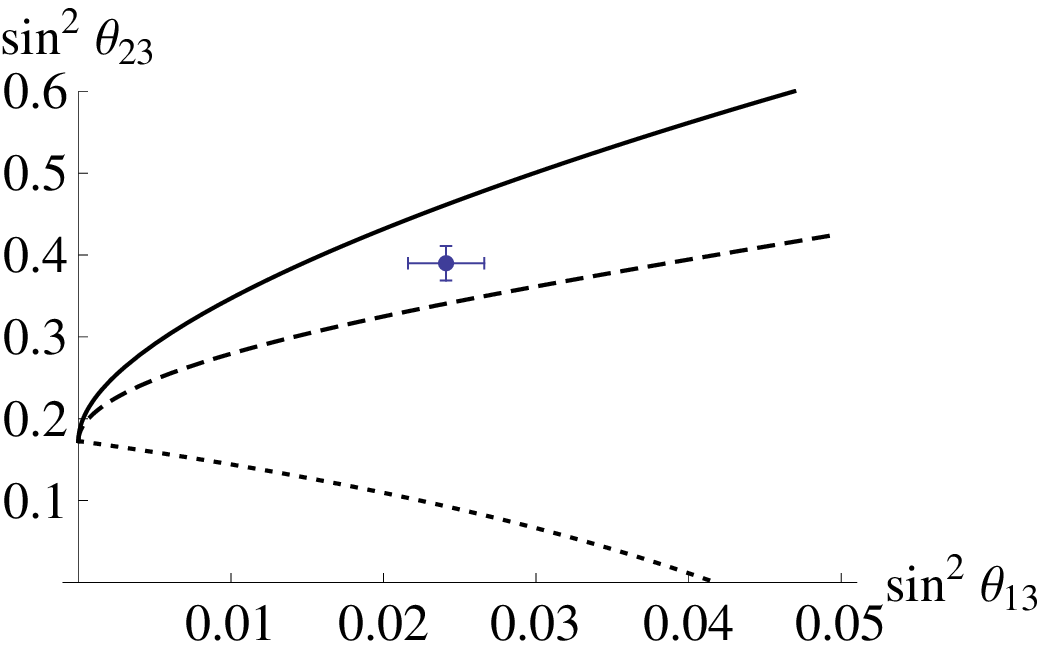}
\caption{
$\sin^2\theta_{12}$ (left panels) and  $\sin^2\theta_{23}$ (right panels) as  functions of $\sin^2\theta_{13}$ for the patterns \emph{PSL7A} (top panels) and \emph{PSL7B} (bottom panels). The curves  in the right panel correspond to $\delta = 0$ (thick), $\pi/4$ (dashed),  $\pi/2$ (dotted).}
\label{newfig2}
\end{center}
\end{figure}

\begin{figure}[ht]
\begin{center}
\includegraphics[width=7.5cm]{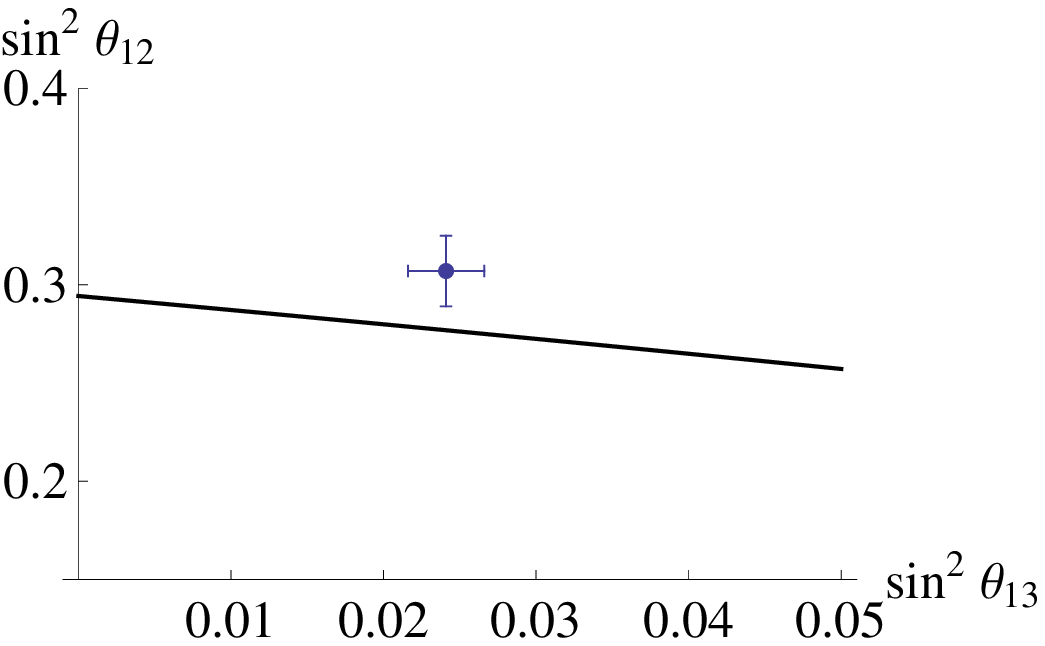}
\includegraphics[width=7.5cm]{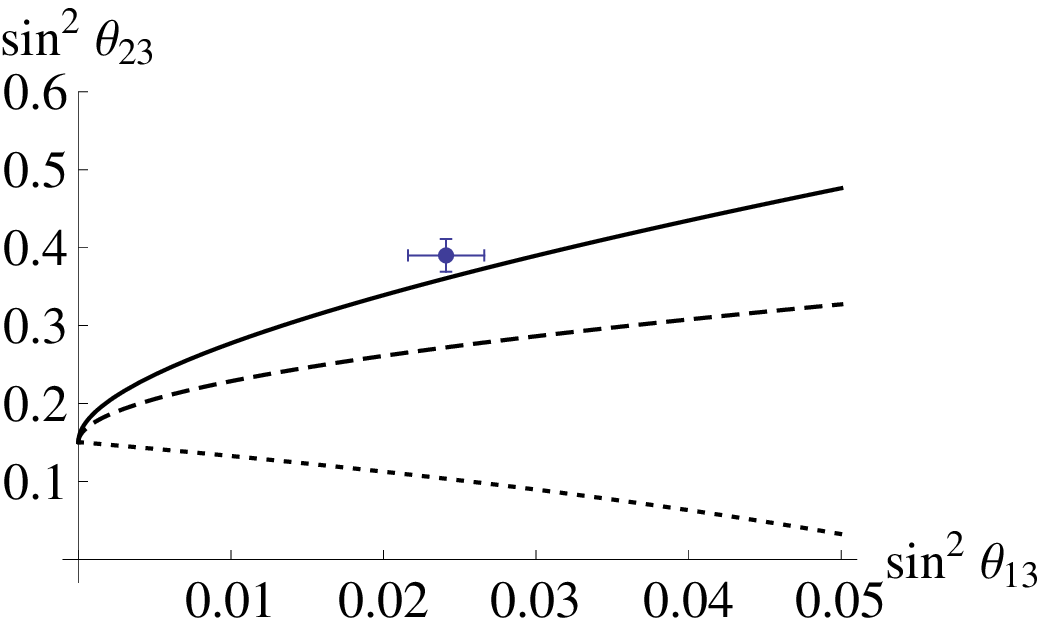}
\caption{
$\sin^2\theta_{12}$ (left panel) and  $\sin^2\theta_{23}$ (right panel) as  functions of $\sin^2\theta_{13}$ for the pattern \emph{PSL7C}. The curves in the right panel correspond to $\delta = 0$ (thick), $\pi/4$ (dashed),  $\pi/2$ (dotted).}
\label{newfig2a}
\end{center}
\end{figure}

Since $q = 4$, the eigenvalues of $X$ are 4th roots of unity and can be taken as  $\{1,\,i,\,-i\}$,  for which we have $x = 1$.  
For $p = 3$ we have the equality in Eq.~(\ref{trXp=3}) which gives $A = 1$. Substituting this value into Eq.~(\ref{a-phases}) we find that it is satisfied for $k_e = 5$, $k_\mu = 3$ and $k_\tau = 6$.  Thus, the symmetry assignment is
\be
\{ m,\, k_e,\, k_\mu,\,p,\,a \} = \{ 7, 5, 3, 3, 0\}.
\label{set7}
\ee
We can now plug in the values of $k_e$ and $k_\mu$ in Eqs.~(\ref{Uei}-\ref{Utaui}) 
to obtain the values of the mixing elements: 
\be
|U_{\mu i}|^2 = \frac{1}{4\left[ 1 + 
\sin \frac{\pi}{14} \right]} \,,\quad |U_{\tau i}|^2 = \frac{1}{4 \left[ 1 + \cos \frac{\pi}{7} \right]},  \label{Umu1PSL7} 
\ee
and 
$|U_{ei}|^2 = 1 - |U_{\mu i}|^2 -  |U_{\tau i}|^2$. 

For  $i = 2$, the values of mixing parameters in Eq.~(\ref{Umu1PSL7}) are in a good agreement with experiment. We call this mixing pattern, the \emph{PSL7A} pattern - see upper panels of Fig.~\ref{newfig2}.  The predicted CP phase has a central value of $\delta = 87.5^\circ$ and for the 1$\sigma$ allowed regions of mixing angles, $\delta$ in the interval $\delta = 82^\circ - 93^\circ$.

Note that $T$ is fixed up to permutations of its eigenvalues. The predicted entries of $U_{PMNS}$ depend not only on $i$ but also on the particular ordering of the $k_\alpha$ in $T$. For \emph{PSL7A}, we chose the order which gives the best agreement with experimental data.

\vtw
Consider now 
$mp$-permuted assignment 
\be
\{ m, ~k_e', ~k_\mu',\,p,\,a \} = \{ 3,~1, ~ 0,\,7, ~ - (1+ i\sqrt{7})/2 \} .
\ee
Now the mixing pattern can be found directly from the mixing obtained in the previous case as described in Sec.~\ref{sec2}. We get:  
\be
\mcl{V}_i^T =  \left[ \frac{1}{12}\left( 5 + \sqrt{21}\right ), \,  
\frac{1}{6}, \,  \frac{1}{12}\left( 5 - \sqrt{21}\right) \right] \,.
\ee
For $i = 1$, we call this pattern \emph{PSL7B}. The values of $\sin^2\theta_{12}$ and $\sin^2\theta_{23}$ as functions of $\sin^2\theta_{13}$ are shown in the bottom panels of Fig.~\ref{newfig2}. The predicted value of $\theta_{12}$ is substantially smaller than the observed one in this case. Thus, large corrections would be required in order to make the pattern viable.

\vth
Besides Eq.~(\ref{prepsl}) the group $PSL(2,\,\mbf{Z}_7)$ has the presentation
\be
S_{iU}^2 = T^7 = (S_{iU}T)^4 = \mbb{I} \,,\quad X^4 = \mbb{I} . 
\label{psl7a}
\ee
It leads to the same values for the mixing parameters as in Eq.~(\ref{Umu1PSL7})  if $\{k_e ,\, k_\mu,\, k_\tau\} = \{6,\, 5,\, 3\}$, {\it i.e.}, if the $T$-charges are permuted with respect to those in Eq.~(\ref{set7}).

A new mixing pattern appears in the case of $mp$-permuted assignment:
\be
\{ m, ~k_e', ~k_\mu',\,p, ~a'\} = \{ 4, ~3, ~ 0,\,7, ~ - (1+ i\sqrt{7})/2 \}.
\ee
In this case we find 
\be
\mcl{V}_i^T = \left(\frac{1}{8}\left( 3 + \sqrt{7} \right), \, 
\frac{1}{4},  \, \frac{1}{8}\left( 3 - \sqrt{7} \right) \right). 
\label{psl7c}
\ee
For $i=1$, the column in Eq.~\eqref{psl7c}, which we refer as pattern \emph{PSL7C}, leads to the mixing angles shown in Fig.~\ref{newfig2a}. Although this case  still requires some corrections, the disagreement with data is smaller that in the pattern \emph{PSL7B}.


\subsection{Symmetry relations from \large{$\Delta(96)$}}


The group $\Delta(96)$ is a finite subgroup of the infinite von Dyck group with presentation
\be
S_{iU}^2 = T^8 = (S_{iU}T)^3 = \mbb{I} \, 
\ee
\begin{figure}[t]
\begin{center}
\includegraphics[width=7.5cm]{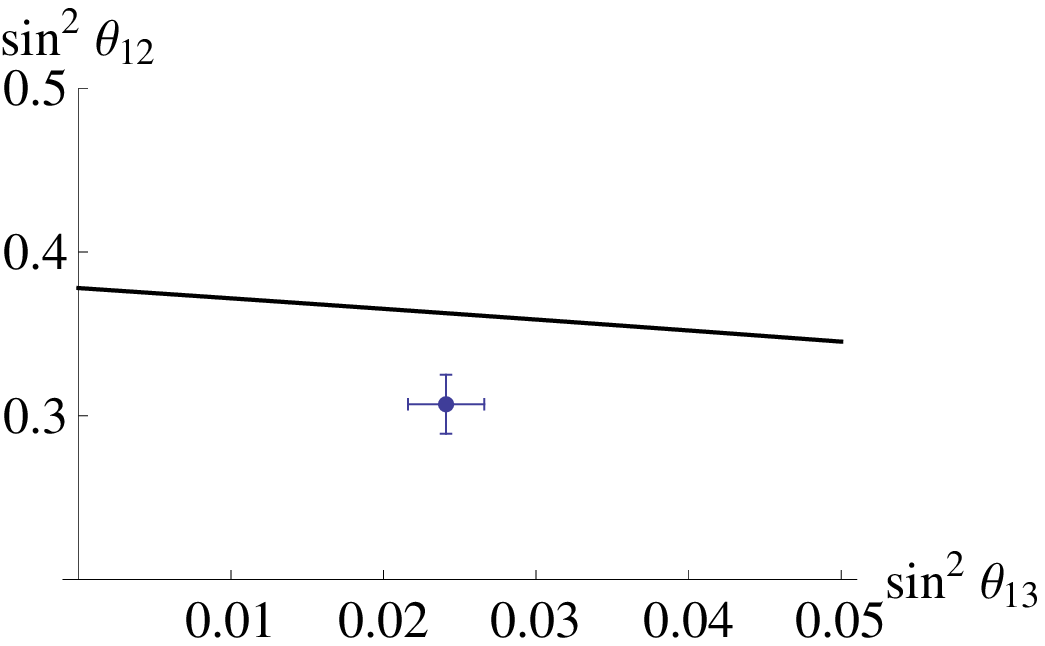}
\includegraphics[width=7.5cm]{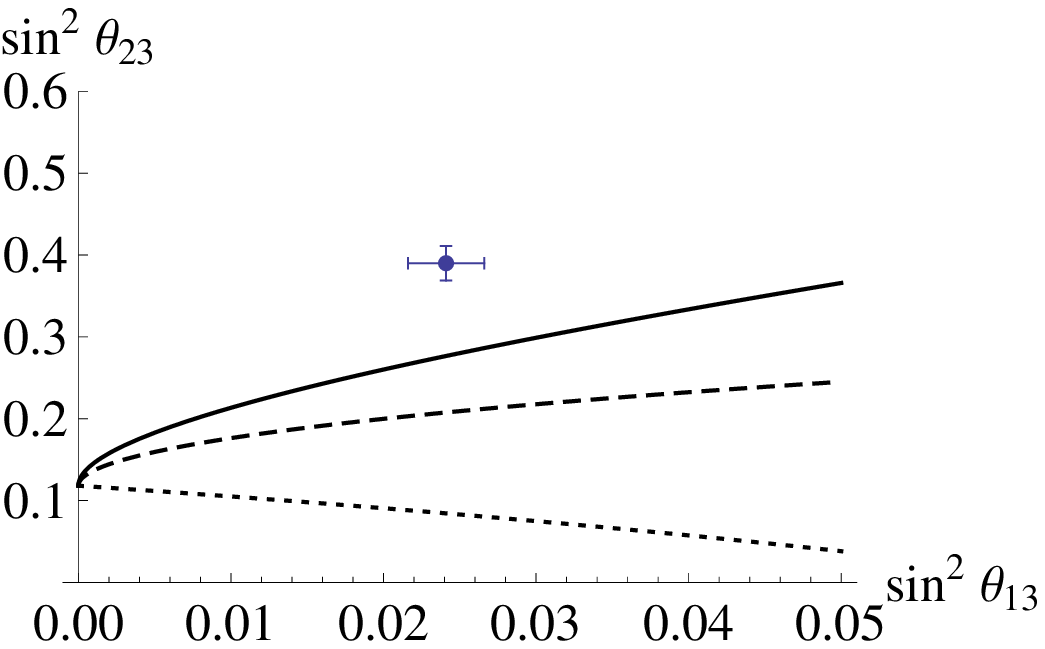}
\caption{
$\sin^2\theta_{12}$ (left panel) and  $\sin^2\theta_{23}$ (right panel) as  functions of $\sin^2\theta_{13}$ for the pattern \emph{D96}. The curves  in the right panel correspond to $\delta = 0$ (thick), $pi/4$ (dashed),  $\pi/2$ (dotted).}
\label{newfig3}
\end{center}
\end{figure}
to which the additional relation
\be
X^3 = \mbb{I} \label{rel5}
\ee
is added in order to render it finite.
From Eq.~\eqref{rel5} one obtains $x = 0$ and, consequently, we must have $A = 0$ according to Eq.~(\ref{trXp=3}) which holds for $p = 3$. For vanishing $A$, Eq. (\ref{a-phases}) is satisfied for $k_e = 5$ and $k_\mu = 2$. Substituting these values in Eqs.~(\ref{Uei}-\ref{Utaui}) we find the column $\mcl{V}_i^T =  \{1/4,\, 1/4,\, 1/2\}$ and the possibilities that derive from all possible permutations. These cases were studied in Sec.~\ref{sec2}. 

Turning to the $mp$-permuted assignment, we have
\be
\{ m, ~k_e', ~k_\mu',\,p, ~a'\} = \{ 3,  ~1, ~ 0,\,8, ~i \}
\ee 
which leads to the mixing pattern
\be
\mcl{V}_i^T = \left( \frac{2+\sqrt{3}}{6},\, \, \frac{1}{3},\, \frac{2-\sqrt{3}}{6} \right).
\ee
The mixing angles which follow from this column setting $i=1$, mixing pattern \emph{D96}, are shown in Fig.~\ref{newfig3}.  There is no  good agreement with observations and large corrections are required for this pattern to be viable.



\subsection{Symmetry relations from \large{$\Delta(384)$}}


The group $\Delta(384)$ is the subgroup of the infinite von Dyck group with presentation
\be
S_{iU}^2 = T^{16} = (S_{iU}T)^3 = \mbb{I} \,,
\ee
where $S_{iU}$ and $T$ satisfy the additional relation
\be
X^3 = \mbb{I}.  
\label{rel6}
\ee
Now, $a = x = 0$ and therefore $A = 0$. Due to large $m = 16$ there are many a priori possibilities for the matrix $T$. For zero $A$ and $m = 16$ we find that Eq.~(\ref{a-phases}) is satisfied for the two sets of values of $\{k_e,\,k_\mu\}$:  $\{-7,\,1\}$ and  $\{-5, \,3\}$.
\begin{figure}[ht]
\begin{center}
\includegraphics[width=7.5cm]{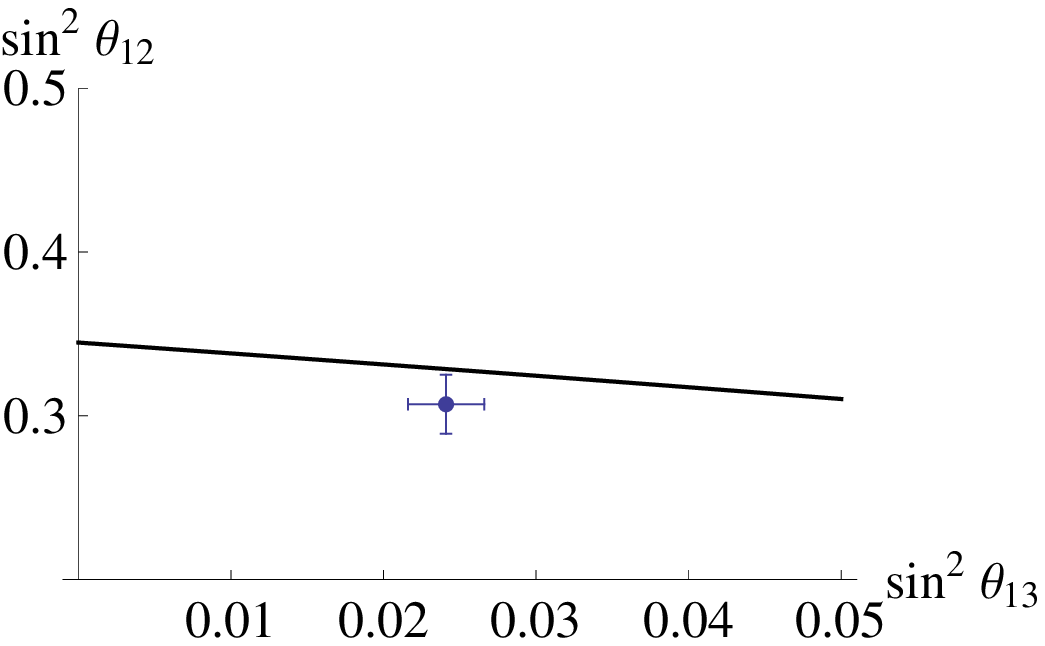}
\includegraphics[width=7.5cm]{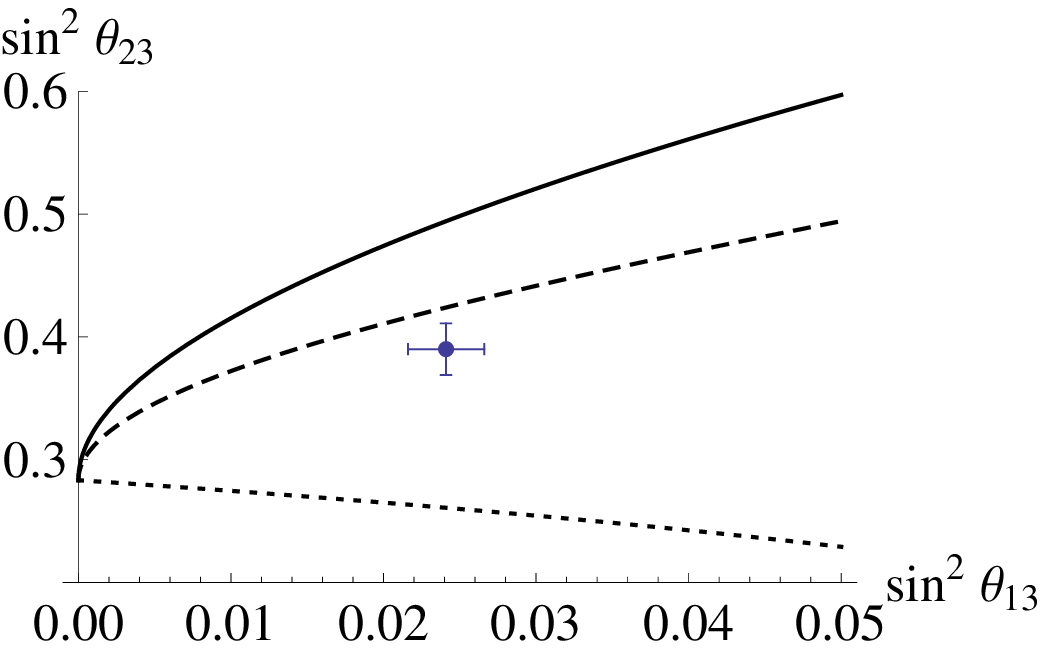}
\caption{
$\sin^2\theta_{12}$ (left panel) and  $\sin^2\theta_{23}$ (right panel) as  functions of $\sin^2\theta_{13}$ for the pattern  \emph{D384A}. The curves  in the right panel correspond to $\delta = 0$ (thick), $\pi/4$ (dashed),  $\pi/2$ (dotted).}
\label{newfig4}
\end{center}
\end{figure}
No new mixing pattern appears when $T$ is the residual symmetry of the charged lepton sector. However, the $mp$-permuted assignment leads to 
interesting  mixing patterns which correspond to two possible choices of $a'$. 
In the first case, for $\{k_e,\,k_\mu\} = \{-7,\,1\}$ one finds the assignment  
\be
\{ m, ~k_e', ~k_\mu',\,p, ~a'\} = \left\{ 3, ~1, ~1 ,\,16, \frac{-1+i}{\sqrt{2}} \right\} 
\ee 
and then 
\be
\mcl{V}_i^T = \left( \frac{4+\sqrt{2}+\sqrt{6}}{12},\, \, \frac{4+\sqrt{2}-\sqrt{6}}{12},
\, \, \frac{2-\sqrt{2}}{6} 
\right). 
\ee
For $i=1$ we call  this pattern \emph{D384A} and  the predicted mixing angles are plotted in Fig.~\ref{newfig4}. This pattern is able to fit all mixing angles for $\delta = 55^\circ$  with the $1\sigma$ range for the mixing angles giving a range for $\delta$, $49^\circ < \delta < 61^\circ$. 

In the second case, $\{k_e,\,k_\mu\} = \{-5, \,3\}$, the $mp$-permuted assignment is
\be
(m, ~k_e', ~k_\mu',\,p, ~a') = \left(3, ~ - 1, ~1,\,16, \frac{1+i}{\sqrt{2}}, \right).
\ee
It leads to 
\be
\mcl{V}_i^T  = \left( \frac{4+\sqrt{2}+\sqrt{6}}{12},\, 
\frac{4+\sqrt{2}-\sqrt{6}}{12},\, \frac{2-\sqrt{2}}{6} \right). 
\ee
This column comes closer to the experimental values for $i=3$. In that case, we call this pattern \emph{D384B}.  It leaves $\delta$ and $\theta_{12}$ undetermined but gives predictions for $\theta_{13}$ and $\theta_{23}$:  
\be
\sin^2\theta_{12} = \frac{4-\sqrt{2}-\sqrt{6}}{12} \approx 0.011\,,\quad 
\sin^2\theta_{23} = \frac{4-\sqrt{2}+\sqrt{6}}{8+\sqrt{2}+\sqrt{6}} \approx 0.424 \,.
\ee
in very good agreement with experiment. 


\section{The residual Klein group for neutrinos}
\label{sec6}


\vth
Let us take the full Klein group $\mbf{Z}_2\otimes \mbf{Z}_2 $ as the residual symmetry of the neutrino sector. Such a possibility has been realized in most of the specific models \cite{A4TBMmodels, S4BMmodels, S4TBMmodels}. For definiteness,  we take  $S_{1U}$ and $S_{2U}$ as the generators of the group, so that its  presentation is given by
\footnote{Which $S_{iU}$ are chosen in Eq.~\eqref{kleing-rels} 
is irrelevant due to Eq.~\eqref{S3U}.}
\be
S_{1U}^2 = S_{2U}^2 = T^m = W_{1U}^{p_1} = W_{2U}^{p_2} = \mbb{I} \,,\quad [S_{1U},\,S_{2U}] = 0 .  
\label{kleing-rels}
\ee
In Eq.~\eqref{kleing-rels} we have two sets of relations, one for $S_{1U}$ and another for $S_{2U}$, of the type considered in the previous sections.
These relations impose two sets of conditions on $U_{PMNS}$ which can be written as
\be
\sum_{\alpha} e^{i\phi_\alpha}\xi_{\alpha i} = a_i \,,\quad \trm{for } i = 1,\,2 ,  
\label{ai}
\ee
where $a_i \equiv \sum_j \lambda^{(p_i)}_j$ is a sum of three $p_i$-th roots of unity. 
Eq.(\ref{ai}) gives four relations, that determine two columns of the mixing matrix, which are enough to determine $U_{PMNS}$ completely. An example of this situation appears in Eqs.~(\ref{TM1}) and (\ref{TM2}) with $\{p_1,\,p_2,\, m\} = \{4,\,3, \, 3\}$. The only mixing matrix compatible with both  Eqs.~(\ref{TM1}) and (\ref{TM2}) is the TBM.

\vth
In \cite{toorop} it was found that, in some cases, the mixing matrix derived from Eq.~\eqref{ai} has the property that the absolute values of the entries of two column were equal up to a permutation.  In what follows we  provide a general explanation of this fact.
 

Consider the elements $g_k$ of $G_f$ that can be written as
\be
g_k = S_{1U}T^k 
\label{ST^k}, 
\ee
where $k$ is an integer.  
If $G_f$ is finite then there exists an integer $r$, with $r < m$, such that  
\be
g^{r}_k =  (S_{1U}T^k)^r =  \mbb{I} \,.
\label{grrel}
\ee
Notice that $g_k$ generates a  $\mbf{Z}_r$ subgroup of $G_f$.

 Eq.~(\ref{grrel}) implies that that the entries of $U_ {PMNS}$ satisfy another relation, similar to Eq.~\eqref{ai} for i=1, but extracted by substituting $T \rightarrow T^k$. This new relation can be obtained from Eq.~\eqref{ai} by replacing $a$ and the  phases correspondingly. We obtain
\be
\sum_{\alpha} e^{i k \phi_\alpha} \xi_{\alpha 1} =  \tilde{a}_k(r) \, ,  
\label{trgk}
\ee
where $\tilde{a}_k(r)  \equiv \sum \lambda_j^{(r)}$.
Suppose now that there exists a value of $k$ such that 
\begin{align}
 \{ k \phi_e\,, k \phi_\mu,\, k \phi_\tau\} & = \trm{permutation of } \{\phi_e\,,\phi_\mu,\,\phi_\tau\} \mod 
2\pi 
\nonumber\\
&\equiv \{\phi_{h(e)}\,,\phi_{h(\mu)},\,\phi_{h(\tau)}\} \, ,  
\label{kperm}
\end{align}
where $h$ represents a permutation of $e$, $\mu$ and $\tau$. In that case, Eqs. (\ref{trgk}) can be expressed as 
\be
\sum_{\alpha} e^{i \phi_{h(\alpha)}} \xi_{\alpha 1} = \tilde{a}_k(r) \,   
\ee
and, according to Eq.~(\ref{kperm}), this can be rewritten as 
\be
\sum_{\alpha} e^{i \phi_{\alpha}} \xi_{f(\alpha) 1} = \tilde{a}_k(r) \, ,  
\label{hatak}
\ee
where $f \equiv h^{-1}$ is the inverse permutation of $h$. If moreover
\be
\tilde{a}_k(r) =   a_2 \, , 
\label{a2=ak}
\ee
then Eq.~\eqref{hatak} becomes
\be
\sum_{\alpha} e^{i \phi_{\alpha}} \xi_{f(\alpha) 1} = a_2 .   
\label{hatak2}
\ee
Confronting this equation with Eq. (\ref{ai}) for $i = 2$  we conclude that 
\be
\xi_{\alpha 2}  =  \xi_{f(\alpha) 1} \,.
\label{perm}
\ee
That is, the elements of column 2 are equal to the  permuted elements of  column 1.  

Eqs.~\eqref{kperm} and \eqref{a2=ak} are the necessary and sufficient conditions for the second column of $U_{PMNS}$ to be a permutation of the first one. The analysis does not depend on which $i$ and $j$ are chosen in Eq.~\eqref{kleing-rels}, so that the same arguments apply to other pairs of columns of $U_{PMNS}$.

\vth
We illustrate this permutation  property with the groups studied in Sec.~5. 
The pattern \emph{PSL7A} corresponds to  $T$ with phases 
\be
\{ \phi_\alpha \} \equiv \{ \phi_e, \phi_\mu,  \phi_\tau \} = 
\frac{2\pi}{7} \{5, \, 3, \, 6\}. 
\ee
Considering the group element $g_2 = S_{1U}T^2$,  
so that  $k=2$ in Eq.~\eqref{ST^k},  we find that 
\be
2\{ \phi_\alpha \} \mod 2\pi =  \frac{2\pi}{7} \{3, \, 6, \, 5\}. 
\ee
Therefore, according to Eq.~(\ref{perm}), we have 
\be
\xi_{e j} = \xi_{\tau 1}, ~~\xi_{\mu j} = \xi_{e1}, ~~ \xi_{\tau j} = \xi_{\mu 1}   
\label{perxi}
\ee
provided that 
\be
a_j = \tilde{a}_2 \,. 
\label{perma}
\ee
The elements $|U_{\alpha j}|^2$ for $j\neq 1$ are a cyclic permutation 
of $|U_{\alpha 1}|^2$. According to Eq.~(\ref{perxi})
for  $j=2$ the full matrix of mixing parameters equals 
\be
||~ |U_{\alpha i}|^2 || = 
\left( \begin{array}{ccc}
c_e & c_\tau & c_\mu \\
c_\mu & c_e & c_\tau  \\
c_\tau & c_\mu & c_e
\end{array} \right) ,  
\label{PSL7fullU}
\ee
where $c_e \equiv |U_{e 1}|^2$,  $c_\mu \equiv |U_{\mu 1}|^2$ and  $c_\tau \equiv |U_{\tau 1}|^2$   are given in Eq.~(\ref{Umu1PSL7}).

Once $U_{PMNS}$ is found, the validity of the assumption in Eq.~\eqref{perma} can be checked. Knowing the explicit expression for $||~ |U_{\alpha i}|^2 ||$, 
Eq.~\eqref{PSL7fullU}, we can calculate the traces of the matrices $S_{1U}T^2$ and $S_{2U}T$:  
$\tr{S_{1U}T^2} = \tr{g_2} = \tilde{a}_2$, $\tr{S_{2U}T} = \tr{W_{U2}} = a_2$
and check that they are equal. 

Note that in the case of pattern \emph{PSL7A} 
we can also choose $k = 4$ in Eq.~\eqref{kperm} which leads to
\be
a_j = \tilde{a}_4(r)  \,,\quad \{\xi_{\tau j},\,\xi_{e j},\,\xi_{\mu j}\} =  \{\xi_{e 1},\,\xi_{\mu 1},\,\xi_{\tau 1}\}  \,. 
\label{perma2}
\ee
This is consistent with the permutation of $|U_{\alpha 1}|^2$ 
that appears in the third column, $|U_{\alpha 3}|^2$, in Eq.~\eqref{PSL7fullU}. 
These columns can, of course, be interchanged. 
Among the various possibilities in the context of \emph{PSL7A}, 
the matrix in Eq.~\eqref{PSL7fullU} is the closest one to the experimentally determined matrix. Therefore  the choice of group parameters and permutations, which leads to Eq.~\eqref{PSL7fullU} is the best choice from the experimental point of view. 

The same method  can be also applied to the patterns \emph{PSL7B}, \emph{PSL7C} and \emph{D96}. They all have $m<p$ and, in particular, the charged-lepton generator is $T'$, with $T'^3 = \mbb{I}$. Thus, the eigenvalues of the generator $T'$ are cubic roots of unity which implies that one can only have $k = 2$ in Eq.~\eqref{trgk}. Following the same steps as above, one can find the following matrices of the moduli of mixing elements squared,   
$||~ |U_{\alpha i}|^2~||$:
\be
\frac{1}{12} \left( \begin{array}{ccc}
5+\sqrt{21} & 2 &  5-\sqrt{21} \\
2 & 8 & 2 \\
5-\sqrt{21} & 2 & 5+\sqrt{21}
\end{array} \right) \,,\; \, \,  \frac{1}{8} \left( \begin{array}{ccc}
3+\sqrt{7} & 2 &  3-\sqrt{7} \\
2 & 4 & 2 \\
3-\sqrt{7} & 2 & 3+\sqrt{7}
\end{array} \right)
\ee
for the patterns  \emph{PSL7B}, \emph{PSL7C} correspondingly, and 
\be
\frac{1}{6} \left( \begin{array}{ccc}
2+\sqrt{3} & 2&  2-\sqrt{3} \\
2 & 2 & 2 \\
2-\sqrt{3} & 2 & 2+\sqrt{3}
\end{array} \right) 
\ee
for the pattern \emph{D96}. In contrast to the cyclic permutations of pattern \emph{PSL7A}, in all these cases we find permutations of two elements in  two different columns. This is related to the fact that $T'$ for these models has an eigenvalue that is equal to 1.


\section{Conclusions} 
\label{discussion}


In this paper we extended the formalism presented in \cite{first-paper} in several directions.

\begin{itemize}

\item 
A closed formula for the column elements of the mixing matrix was obtained without any of the assumptions in \cite{first-paper}. This formula allowed us to explore new flavor groups that were not considered in our previous work. 

\item We systematically analized all possible mixing patterns that can be obtained with the finite non-dihedral von Dyck groups. We conclude that  relations obtained in \cite{first-paper} and in Sec. \ref{sec4} of this paper cover all physically relevant relations that can be obtained within this framework. In Sec. \ref{sec4} we present a mixing pattern so far overlooked in the literature in which the elements of the second column are predicted by the symmetry to be  $\{1/4,\,1/2,\,1/4\} $. It corresponds to a transposition of the BM mixing pattern. This mixing pattern does not provide a good fit but could be used as a zero order pattern for which significant corrections are required. 

\item The groups  $PSL(2,\mbf{Z}_7)$, $\Delta(96)$ and $\Delta(384)$ are subgroups of infinite von Dyck groups and can be analized within our formalism. 
We obtained the constraints on $U_{PMNS}$ for these cases and showed that the symmetry assignments corresponding to the patterns \emph{PSL7A} and \emph{D384A} provide an excellent agreement with experiment. For these patterns we have predictions for the CP phase: $\delta \sim 87^\circ$ for \emph{PSL7A} and $\delta \sim 55^\circ$ for \emph{D384A}.

\item Imposing the  Klein symmetry in the neutrino sector fixes completely the mixing matrix. Using our approach, we were able to explain the appearance of mixing patterns in which two or three columns of $U_{PMNS}$  are composed of the same but permuted elements. 
\end{itemize}


\section*{Acknowledgements}


D. H. has benefitted from discussions with Claudia Hagedorn and Luca Merlo and thanks Ferruccio Feruglio for patiently explaining to him different aspects of the theory of discrete groups.

\end{document}